\documentclass{article}
\usepackage[preprint]{spconf,amsmath,graphicx}

\copyrightnotice{978-1-7281-0306-8/19/\$31.00~\copyright2019 IEEE - ASRU 2019}

\def\vecx{{\mbox{\bf x}}}

\def\funcs{{\bf s}}
\def\funct{{\bf t}}

\title{A DENSITY RATIO APPROACH TO LANGUAGE MODEL FUSION \\ IN END-TO-END AUTOMATIC SPEECH RECOGNITION}
\name{Erik McDermott, Hasim Sak, Ehsan Variani}
\address{Google Inc., USA\\
    {\small \tt \{erikmcd,hasim,variani\}@google.com}}

\begin{document}

\maketitle

\begin{abstract}
This article describes a density ratio approach to integrating
external Language Models (LMs) into end-to-end models for Automatic
Speech Recognition (ASR). Applied to a Recurrent Neural Network
Transducer (RNN-T) ASR model trained on a given domain, a matched
in-domain RNN-LM, and a target domain RNN-LM, the proposed method uses
Bayes' Rule to define RNN-T posteriors for the target domain, in a
manner directly analogous to the classic hybrid model for ASR based on
Deep Neural Networks (DNNs) or LSTMs in the Hidden Markov Model (HMM)
framework (Bourlard \& Morgan, 1994). The proposed approach is
evaluated in cross-domain and limited-data scenarios, for which a
significant amount of target domain text data is used for LM training,
but only limited (or no) \{audio, transcript\} training data pairs are
used to train the RNN-T. Specifically, an RNN-T model trained on
paired audio \& transcript data from YouTube is evaluated for its
ability to generalize to Voice Search data. The Density Ratio method
was found to consistently outperform the dominant approach to LM and
end-to-end ASR integration, Shallow Fusion.
\end{abstract}


\section{Introduction}
\label{sec:intro}

End-to-end models such as Listen, Attend \& Spell (LAS)
\cite{ChanIcassp2016} or the Recurrent Neural Network Transducer
(RNN-T) \cite{Graves2012} are sequence models that directly define
$P(W | X)$, the posterior probability of the word or subword sequence
$W$ given an audio frame sequence $X$, with no chaining of sub-module
probabilities.
State-of-the-art, or near state-of-the-art
results have been reported for these models on challenging tasks
\cite{AudhkhasiIcassp2018,ChiuIcassp2018}.

End-to-end ASR models in essence do not include independently trained
symbols-only or acoustics-only sub-components. As such, they do not
provide a clear role for language models $P(W)$ trained only on
text/transcript data. There are, however, many situations where we
would like to use a separate LM to complement or modify a given ASR
system. In particular, no matter how plentiful the paired \{audio,
transcript\} training data, there are typically orders of magnitude
more text-only data available. There are also many practical
applications of ASR where we wish to adapt the language
model, e.g., biasing the recognition grammar towards a list of
specific words or phrases for a specific context.

\begin{figure}
    \centering
    \includegraphics[width=1.0\linewidth]{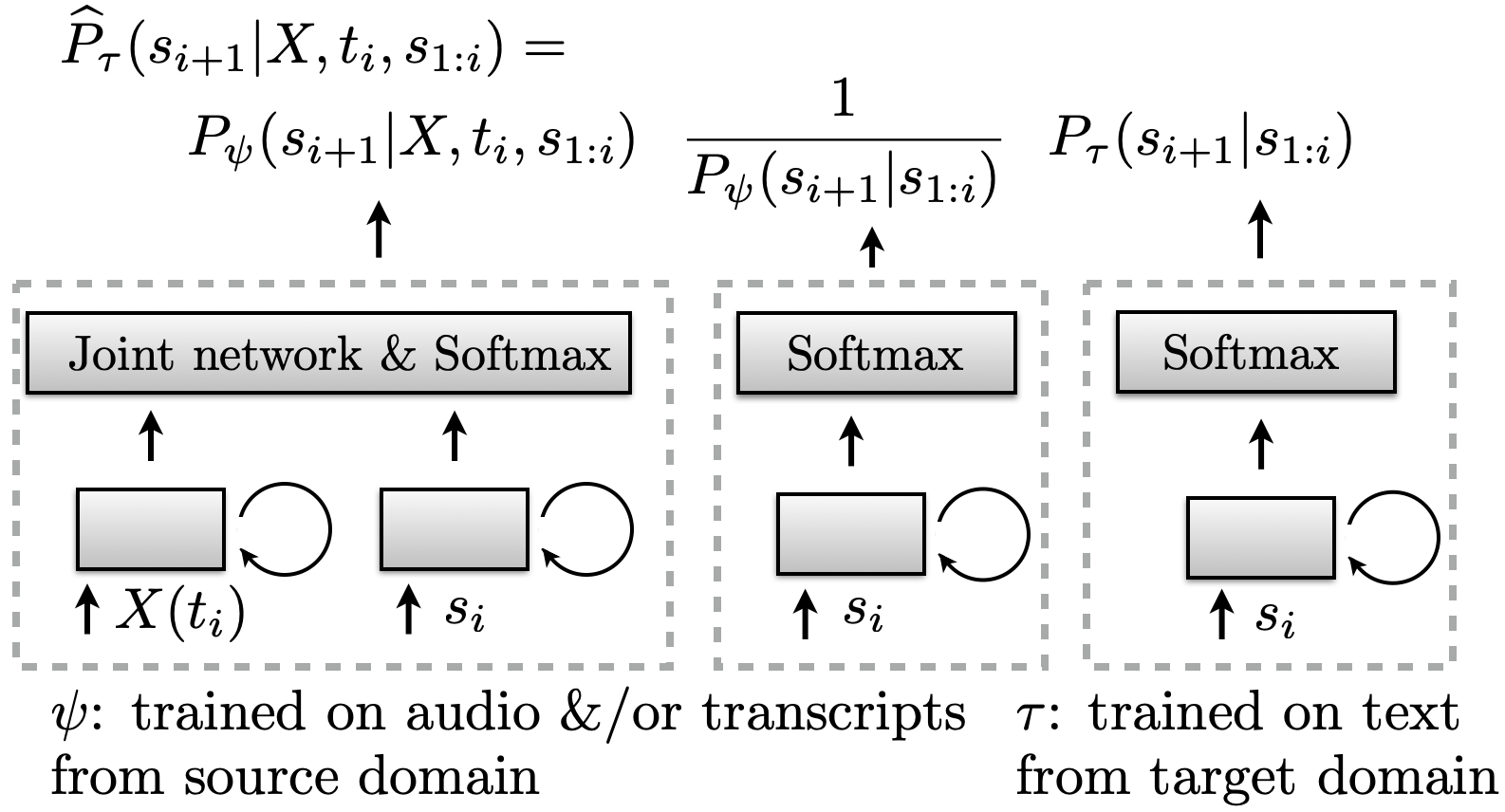}
    \caption{Estimating a target domain pseudo-posterior via combination of
      source domain RNN-T, source domain RNN-LM, and target domain RNN-LM.}
    \label{fig:architecture}
\end{figure}

The research community has been keenly aware of the importance of this
issue, and has responded with a number of approaches,
under the rubric of ``Fusion''. The most popular of these is ``Shallow
Fusion''
\cite{Mikolov2010RecurrentNN,Chorowski2017,hori_asru2017,KannanIcassp2018,toshniwal2018},
which is simple log-linear interpolation between the scores from the
end-to-end model and the separately-trained LM. More structured
approaches, ``Deep Fusion'' \cite{Gulcehre2015}, ``Cold Fusion''
\cite{Sriram2018} and ``Component Fusion'' \cite{ComponentFusion2019}
jointly train an end-to-end model with a pre-trained LM, with the goal
of learning the optimal combination of the two, aided by gating
mechanisms applied to the set of joint scores.
These methods have not replaced the simple Shallow Fusion method as
the go-to method in most of the ASR community.  Part of the appeal of
Shallow Fusion is that it does not require model retraining -- it can
be applied purely at decoding time. The Density Ratio approach
proposed here can be seen as an extension of Shallow Fusion, sharing
some of its simplicity and practicality, but offering a theoretical
grounding in Bayes' rule.

After describing the historical context, theory and practical
implementation of the proposed Density Ratio method, this article
describes experiments comparing the method to Shallow Fusion in a
cross-domain scenario. An RNN-T model was trained on
large-scale speech data with semi-supervised transcripts from YouTube
videos, and then evaluated on data from a live Voice Search service,
using an RNN-LM trained on Voice Search transcripts to try to
boost performance. Then, exploring the transition between cross-domain
and in-domain, limited amounts of Voice Search speech data were used to
fine-tune the YouTube-trained RNN-T model, followed by LM fusion via
both the Density Ratio method and Shallow Fusion. The ratio method
was found to produce consistent gains over Shallow Fusion in all
scenarios examined.

\section{A Brief History of Language Model incorporation in ASR}

{\bf Generative models and Bayes' rule}. The
Noisy Channel Model underlying the origins of statistical ASR
\cite{Jelinek76continuousspeech} used Bayes' rule to combine generative
models of both the acoustics $p(X|W)$ and the symbol sequence $P(W)$:
\begin{eqnarray}
p(X|W) & = & \sum_{\funcs \in S_W} p(X | \funcs) p(\funcs| W) = \sum_{\funcs \in S_W} \prod_t p(\vecx_t | \funcs(t))  \nonumber \\
P(W | X) & = & p(X | W) P(W) / p(X)
\label{equ:channel}
\end{eqnarray}
for an acoustic feature vector sequence $X = \vecx_1, ..., \vecx_T$
and a word or sub-word sequence $W = s_1, ..., s_U$ with possible time
alignments $S_W = \{..., \funcs, ...\}$. ASR decoding then
uses the posterior probability $P(W|X)$.  A prior $p(\funcs | W)$ on
alignments can be implemented e.g. via a simple 1st-order
state transition model. Though lacking in discriminative power, the
paradigm provides a clear theoretical framework for decoupling the
acoustic model (AM) $p(X|W)$ and LM $P(W)$.

{\bf Hybrid model for DNNs/LSTMs within original ASR framework}. The
advent of highly discriminative Deep Neural Networks (DNNs)
\cite{minami90,haffner_eurospeech93,lecun1998gradient,kingsbury_icassp2009,SeideAsru2011}
and Long Short Term Memory models (LSTMs)
\cite{Hochreiter_1997,SakInterspeech2014} posed a challenge to the
original Noisy Channel Model, as they produce phoneme- or state- level
posteriors $P(\funcs(t) | \vecx_t)$, not acoustic likelihoods $p(\vecx_t |
\funcs(t))$. The ``hybrid'' model \cite{Bourlard_KLUWER_1994} proposed the
use of scaled likelihoods, i.e. posteriors divided by separately
estimated state priors $P(w)$.  For bidirectional LSTMs, the
scaled-likelihood over a particular alignment $\funcs$ is taken to be
\begin{eqnarray}
P'(X | \funcs) \equiv k(X) \prod_t P(\funcs(t) | X) / P(\funcs(t)),
\end{eqnarray}
using $k(X)$ to represent a $p(X)$-dependent term shared by all
hypotheses $W$, that does not affect decoding. This
``pseudo-generative'' score can then be plugged into the original
model of Eq.~(\ref{equ:channel}) and used for ASR decoding with an
arbitrary LM $P(W)$.
For much of the ASR community, this approach still constitutes the
state-of-the-art
\cite{AudhkhasiIcassp2018,VarianiInterspeech2018,KaldiPytorchIcassp2019}.

{\bf Shallow Fusion}. The most popular approach to LM incorporation
for end-to-end ASR is a linear interpolation,
\begin{eqnarray}
\text{Score}(X, W) = \log P(W | X) + \lambda \log P(W) + \beta |W|,
\label{equ:shallow_fusion}
\end{eqnarray}
with no claim to direct interpretability according to probability
theory, and often a reward for sequence length $|W|$, scaled by a factor $\beta$
\cite{Chorowski2017,KannanIcassp2018,toshniwal2018,Graves06connectionisttemporal}.

\section{Language Model incorporation into End-to-end ASR, using Bayes' rule}

\subsection{A Sequence-level Hybrid Pseudo-Generative Model}
\label{sec:deep_hybrid}

The model makes the following assumptions: 
\begin{enumerate}
\item The source domain $\psi$ has some true joint distribution $P_{\psi}(W, X)$
over text and audio;
\item The target domain $\tau$ has some other true joint distribution $P_{\tau}(W, X)$;
\item A source domain end-to-end model (e.g. RNN-T) captures $P_{\psi}(W | X)$ reasonably well;
\item Separately trained LMs (e.g. RNN-LMs) capture $P_{\psi}(W)$ and
  $P_{\tau}(W)$ reasonably well;
\item $p_{\psi}(X | W)$ is roughly equal to $p_{\tau}(X | W)$,
  i.e. the two domains are acoustically consistent; and
\item The target domain posterior, $P_{\tau}(W | X)$, is unknown.
\end{enumerate}

The starting point for the proposed {\bf Density Ratio Method} is then to
express a ``hybrid'' scaled acoustic likelihood for the source domain,
in a manner paralleling the original hybrid model
\cite{Bourlard_KLUWER_1994}:
\begin{eqnarray}
p_\psi(X|W) = p_{\psi}(X) P_{\psi}(W|X)/P_{\psi}(W).
\label{equ:hybrid_source}
\end{eqnarray}
Similarly, for the target domain:
\begin{eqnarray}
p_\tau(X|W) = p_{\tau}(X) P_{\tau}(W|X)/P_{\tau}(W).
\label{equ:hybrid_target}
\end{eqnarray}
Given the stated assumptions, one can then estimate the target domain posterior as:
\begin{eqnarray}
\widehat{P}_{\tau}(W | X) = k(X) \frac{P_{\tau}(W)}{P_{\psi}(W)} P_{\psi}(W|X),
\label{equ:updated_target_posterior}
\end{eqnarray}
with $k(X) = p_{\psi}(X) / p_{\tau}(X)$ shared by all hypotheses $W$,
and the ratio $P_{\tau}(W) / {P_{\psi}(W)}$ (really a probablity mass
ratio) giving the proposed method its name.

In essence, this model is just an application of Bayes' rule to
end-to-end models and separate LMs.  The approach can be
viewed as the sequence-level version of the classic hybrid
model \cite{Bourlard_KLUWER_1994}.  Similar use of Bayes' rule to
combine ASR scores with RNN-LMs has been described elsewhere, e.g. in
work connecting grapheme-level outputs with word-level LMs
\cite{hori_asru2017, hori_e2e_rnnlm_2018, Kanda2017}. However, to our knowledge
this approach has not been applied to end-to-end models in
cross-domain settings, where one wishes to leverage a language model
from the target domain. For a perspective on a ``pure'' (non-hybrid)
deep generative approach to ASR, see \cite{McDermott2018}.

\subsection{Top-down fundamentals of RNN-T}

The {\em RNN Transducer (RNN-T)} \cite{Graves2012} defines a
sequence-level posterior $P(W|X)$ for a given acoustic feature vector
sequence $X = \vecx_1, ..., \vecx_T$ and a given word or sub-word
sequence $W = s_1, ..., s_U$ in terms of possible alignments $S_W
= \{..., (\funcs, \funct), ... \}$ of $W$ to $X$. The tuple $(\funcs,
\funct)$ denotes a specific alignment sequence, a symbol sequence and
corresponding sequence of time indices, consistent with the sequence
$W$ and utterance $X$.  The symbols in $\funcs$ are elements of an
expanded symbol space that includes optional, repeatable blank symbols
used to represent acoustics-only path extensions, where the time index
is incremented, but no non-blank symbols are added. Conversely,
non-blank symbols are only added to a partial path time-synchronously.
(I.e., using $i$ to index elements of $\funcs$ and $\funct$,
 $t_{i+1} = t_i + 1$ if $s_{i+1}$ is blank, and $t_{i + 1} = t_i$
if $s_{i+1}$ is non-blank).  $P(W|X)$ is defined by summing over
alignment posteriors:
\begin{eqnarray}
\label{equ:rnnt_sum}
P(W|X) & = & \sum_{(\funcs, \funct) \in S_W} P(\funcs, \funct|X) \\
\label{equ:rnnt}
P(\funcs, \funct|X) & = & \prod_i P(s_{i+1} | X, t_i, s_{1:i}).
\end{eqnarray}
Finally, $P(s_{i+1} | X, t_i, s_{1:i})$ is defined using an LSTM-based
acoustic encoder with input $X$, an LSTM-based label encoder with
non-blank inputs $s$, and a feed-forward joint network combining
outputs from the two encoders to produce predictions for all
symbols $s$, including the blank symbol.

The Forward-Backward algorithm can be used to calculate
Eq.~(\ref{equ:rnnt_sum}) efficiently during training, and Viterbi-based
beam search (based on the argmax over possible alignments) can be used
for decoding when $W$ is unknown \cite{Graves2012,Rabiner93}.

\subsection{Application of Shallow Fusion to RNN-T}

Shallow Fusion (Eq.~(\ref{equ:shallow_fusion})) can be implemented in
RNN-T for each time-synchronous non-blank symbol path extension. The
LM score corresponding to the same symbol extension can be ``fused''
into the log-domain score used for decoding:
\begin{eqnarray}
\text{Score}(s_{i+1} | X, t_i, s_{1:i}) = \log P(s_{i+1} | X, t_i, s_{1:i}) \nonumber \\
+ \lambda \log P(s_{i+1} | s_{1:i}) + \beta.
\end{eqnarray}
This is only done when the hypothesized path extension
$s_{i+1}$ is a non-blank symbol; the decoding score for blank
symbol path extensions is the unmodified $\log P(s_{i+1} | X,
t_i, s_{1:i})$.

\subsection{Application of the Density Ratio Method to RNN-T}

Eq.~(\ref{equ:updated_target_posterior}) can be implemented via an estimated
RNN-T ``pseudo-posterior'', when $s_{i+1}$ is a non-blank symbol:
\begin{eqnarray}
\widehat{P}_{\tau}(s_{i+1} | X, t_i, s_{1:i}) = \frac{P_{\tau}(s_{i+1}|s_{1:i})}{P_{\psi}(s_{i+1}|s_{1:i})} P_{\psi}(s_{i+1}|X, t_i, s_{1:i}).
\label{equ:rnnt_lm_ratio}
\end{eqnarray}
This estimate is not normalized over symbol outputs, but
it plugs into Eq.~(\ref{equ:rnnt}) and Eq.~(\ref{equ:rnnt_sum}) to
implement the RNN-T version of
Eq.~(\ref{equ:updated_target_posterior}).  In practice, scaling
factors $\lambda_\psi$ and $\lambda_\tau$ on the LM scores, and a
non-blank reward $\beta$, are used in the final decoding score:
\begin{eqnarray}
\text{Score}(s_{i+1} | X, t_i, s_{1:i}) = \log P_{\psi}(s_{i+1}|X, t_i, s_{1:i}) \nonumber \\
 + \lambda_\tau \log P_{\tau}(s_{i+1}|s_{1:i}) - \lambda_\psi \log P_{\psi}(s_{i+1}|s_{1:i}) + \beta.
\label{equ:modified_score}
\end{eqnarray}

\subsection{Implementation}
\label{sec:implementation}

The ratio method is very simple to implement. The procedure is essentially to:
\begin{enumerate}
  \item Train an end-to-end model such as RNN-T on a given source
    domain training set $\psi$ (paired audio/transcript data);
  \item Train a neural LM such as RNN-LM on text transcripts from the same training set $\psi$;
  \item Train a second RNN-LM on the target domain $\tau$;
  \item When decoding on the target domain, modify the RNN-T output by
    the ratio of target/training RNN-LMs, as defined in
    Eq.~(\ref{equ:modified_score}), and illustrated in
    Fig.~\ref{fig:architecture}.
\end{enumerate}

The method is purely a decode-time method; no joint training is
involved, but it does require tuning of the LM scaling factor(s) (as
does Shallow Fusion).  A held-out set can be used for that purpose.

\section{Training, development and \\
evaluation data}

\subsection{Training data}
\label{fig:training_data}

The following data sources were used to train the RNN-T and associated RNN-LMs
in this study.

{\bf Source-domain baseline RNN-T}: approximately 120M segmented
utterances (190,000 hours of audio) from YouTube videos, with
associated transcripts obtained from semi-supervised caption filtering
\cite{Liao2013LargeSD}.

{\bf Source-domain normalizing RNN-LM}: transcripts from the same 120M
utterance YouTube training set.  This corresponds to about 3B tokens
of the sub-word units used (see below,
Section~\ref{sec:model_settings}).

{\bf Target-domain RNN-LM}: 21M text-only utterance-level transcripts
from anonymized, manually transcribed audio data,
representative of data from a Voice Search service. 
This corresponds to about 275M sub-word tokens.

{\bf Target-domain RNN-T fine-tuning data}: 10K, 100K, 1M and 21M
utterance-level \{audio, transcript\} pairs taken from anonymized,
transcribed Voice Search data.  These fine-tuning sets roughly
correspond to 10 hours, 100 hours, 1000 hours and 21,000 hours of
audio, respectively.

\begin{table}[ht]
\caption{Training set size and test set perplexity for the morph-level
  RNN-LMs (training domain $\rightarrow$ testing domain) used in this
  study.}
\label{tab:lm_size_ppl}
\begin{center}
\begin{tabular}{|l|c|c|}
\hline
Model & \# Tr. tokens & Test PPL \\
\hline  \hline
YouTube $\rightarrow$ YouTube       & 2.98B & 8.94 \\
\hline
YouTube $\rightarrow$ Voice Search   & 2.98B & 36.5 \\
Voice Search $\rightarrow$ Voice Search   & 275M & 11.1 \\
\hline
\end{tabular}
\end{center}
\end{table}

\subsection{Dev and Eval Sets}

The following data sources were used to choose scaling factors and/or
evaluate the final model performance.

{\bf Source-domain Eval Set (YouTube)}. The in-domain performance of the
YouTube-trained RNN-T baseline was measured on speech data taken from
Preferred Channels on YouTube \cite{SoltauArxiv2016}. The test set is
taken from 296 videos from 13 categories, with each video averaging 5
minutes in length, corresponding to 25 hours of audio and 250,000 word
tokens in total.

{\bf Target-domain Dev \& Eval sets (Voice Search)}. The Voice Search
dev and eval sets each consist of approximately 7,500 anonymized
utterances (about 33,000 words and corresponding to about 8 hours of
audio), distinct from the fine-tuning data described earlier, but
representative of the same Voice Search service. 

\section{Cross-domain evaluation: YouTube-trained RNN-T $\rightarrow$ Voice Search}
\label{sec:yt_to_vs_expts}

The first set of experiments uses an RNN-T model trained on \{audio,
transcript\} pairs taken from segmented YouTube videos, and evaluates
the cross-domain generalization of this model to test utterances taken
from a Voice Search dataset, with and without fusion to an external
LM.

\subsection{RNN-T and RNN-LM model settings}
\label{sec:model_settings}

The overall structure of the models used here is as follows:
\vspace{0.25cm}

{\bf RNN-T}:
\begin{itemize}
\item Acoustic features: 768-dimensional feature vectors obtained from
  3 stacked 256-dimensional logmel feature vectors, extracted every 20
  msec from 16 kHz waveforms, and sub-sampled with a stride of 3, for
  an effective final feature vector step size of 60 msec.
\item Acoustic encoder: 6 LSTM layers x (2048 units with 1024-dimensional
  projection); bidirectional.
\item Label encoder (aka ``decoder'' in end-to-end ASR jargon): 1 LSTM layer
  x (2048 units with 1024-dimensional projection).
\item RNN-T joint network hidden dimension size: 1024.
\item Output classes: 10,000 sub-word ``morph'' units
  \cite{Morfessor} , input via a 512-dimensional embedding.
\item Total number of parameters: approximately 340M 
\end{itemize}

{\bf RNN-LMs} for both source and target domains were set to match the
RNN-T decoder structure and size:
\begin{itemize}
\item 1 layer x (2048 units with 1024-dimensional projection).
\item Output classes: 10,000 morphs (same as the RNN-T).
\item Total number of parameters: approximately 30M.
\end{itemize}

The RNN-T and the RNN-LMs were independently trained on 128-core
tensor processing units (TPUs) using full unrolling and an effective
batch size of 4096. All models were trained using the Adam
optimization method \cite{Adam2015} for 100K-125K steps, corresponding
to about 4 passes over the 120M utterance YouTube training set, and 20
passes over the 21M utterance Voice Search training set.  The trained
RNN-LM perplexities (shown in Table~\ref{tab:lm_size_ppl})
show the benefit to Voice Search test perplexity of training on Voice Search
transcripts.

\begin{figure}
    \centering
    \includegraphics[width=1.0\linewidth]{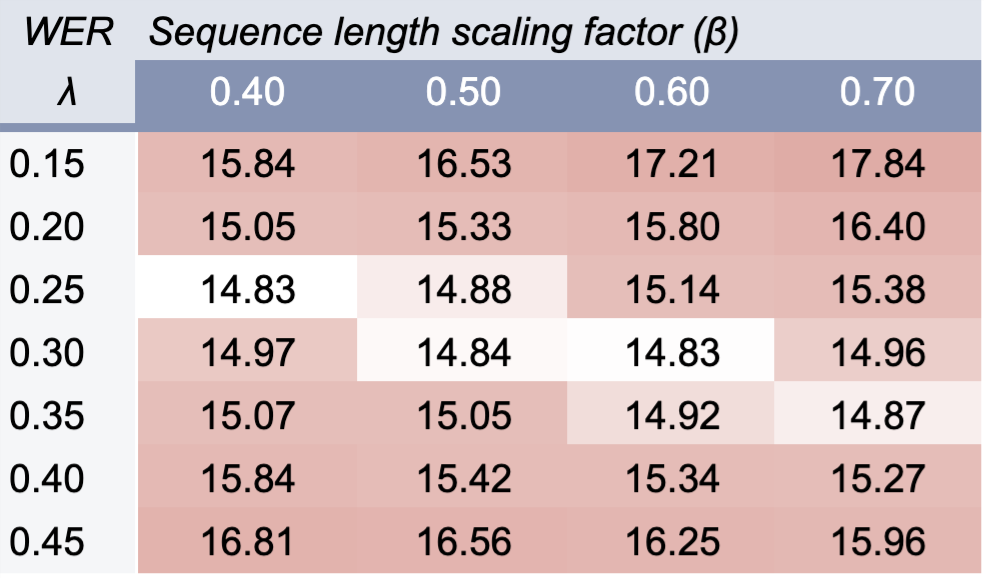}
    \caption{Dev set WERs for Shallow Fusion LM scaling factor $\lambda$ vs.
      sequence length scaling factor $\beta$.}
    \label{fig:heatmap_shallow}
\end{figure}

\begin{figure}
    \centering
    \includegraphics[width=1.0\linewidth]{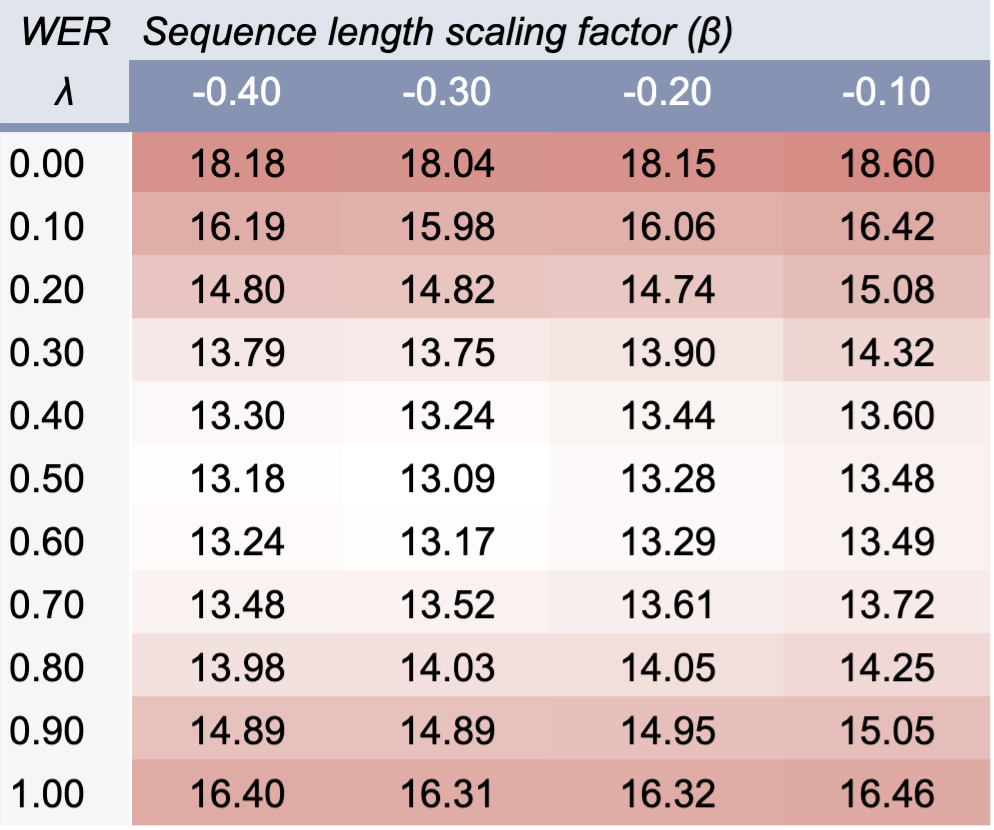}
    \caption{Dev set WERs for Density Ratio LM scaling factor
      $\lambda$ vs. sequence length scaling factor $\beta$. Here
      $\lambda = \lambda_\psi = \lambda_\tau$.}
    \label{fig:heatmap_ratio}
\end{figure}

\subsection{Experiments and results}

In the first set of experiments, the constraint $\lambda_\psi =
\lambda_\tau$ was used to simplify the search for the LM scaling
factor in Eq.~\ref{equ:modified_score}. Fig.~\ref{fig:heatmap_shallow}
and Fig.~\ref{fig:heatmap_ratio} illustrate the different relative
sensitivities of WER to the LM scaling factor(s) for Shallow Fusion
and the Density Ratio method, as well as the effect of the RNN-T
sequence length scaling factor, measured on the dev set.

The LM scaling factor affects the relative value of the symbols-only
LM score vs. that of the acoustics-aware RNN-T score.  This typically
alters the balance of insertion vs. deletion errors. In turn, this
effect can be offset (or amplified) by the sequence length scaling
factor $\beta$ in Eq.~(\ref{equ:shallow_fusion}), in the case of
RNN-T, implemented as a non-blank symbol emission reward.  (The blank
symbol only consumes acoustic frames, not LM symbols
\cite{Graves2012}). Given that both factors have related effects
on overall WER, the LM scaling factor(s) and the sequence length
scaling factor need to be tuned jointly.

Fig.~\ref{fig:heatmap_shallow} and Fig.~\ref{fig:heatmap_ratio}
illustrate the different relative sensitivities of WER to these
factors for Shallow Fusion and the Density Ratio method, measured on
the dev set. 

\begin{figure}
    \centering
    \includegraphics[width=1.0\linewidth]{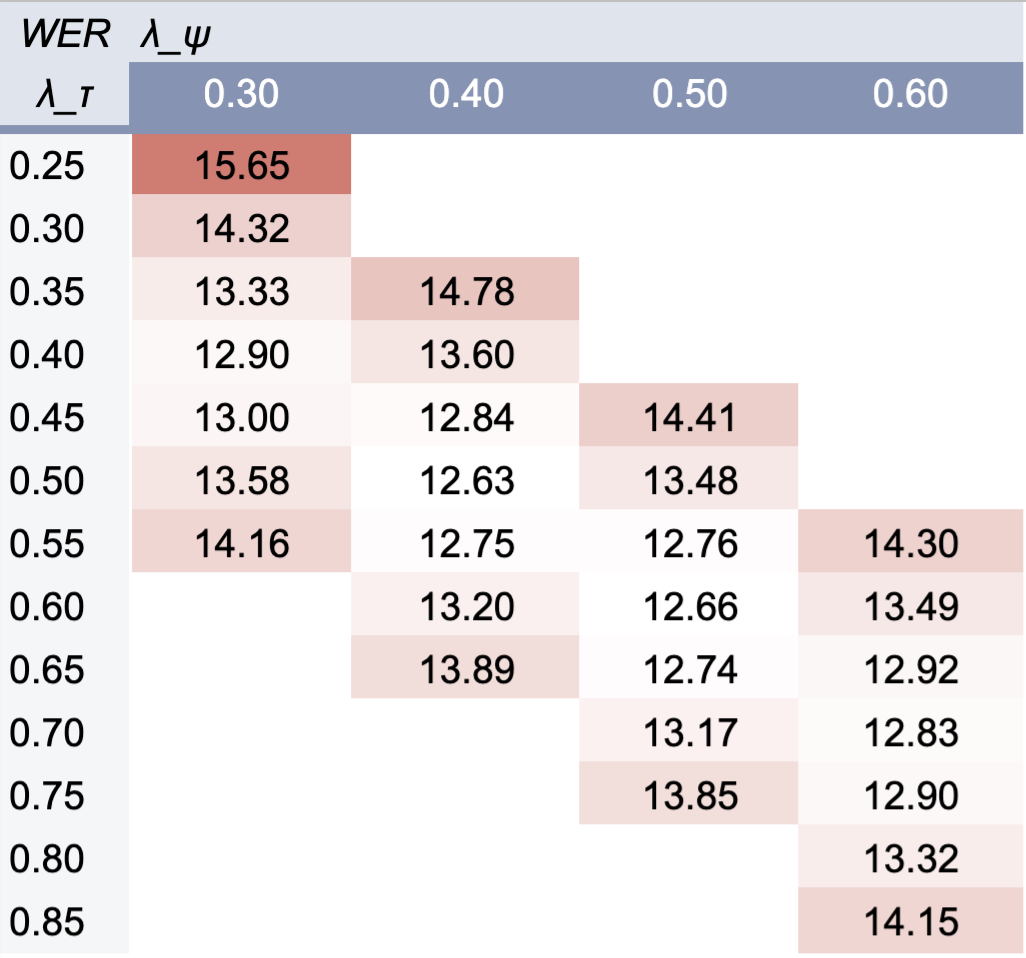}
    \caption{Dev set WERs for different combinations of $\lambda_\tau$
      and $\lambda_\psi$; sequence length scaling factor $\beta=
      -0.1$}.
    \label{fig:heatmap_ratio_lambdas}
\end{figure}

In the second set of experiments, $\beta$ was fixed at -0.1, but the
constraint $\lambda_\psi = \lambda_\tau$ was lifted, and a range of
combinations was evaluated on the dev set.  The results are shown in
Fig.~\ref{fig:heatmap_ratio_lambdas}. The shading in
Figs.~\ref{fig:heatmap_shallow}, \ref{fig:heatmap_ratio} and
\ref{fig:heatmap_ratio_lambdas} uses the same midpoint value of 15.0
to highlight the results.

The best combinations of scaling factors from the dev set evaluations
(see Fig.~\ref{fig:heatmap_shallow}, Fig.~\ref{fig:heatmap_ratio} and
Fig.~\ref{fig:heatmap_ratio_lambdas}) were used to generate the final
eval set results, WERs and associated deletion, insertion and
substitution rates, shown in
Table~\ref{tab:results_morphemes_vs}. These results are summarized in
Table~\ref{tab:results_morphemes_vs_ft}, this time showing the exact
values of LM scaling factor(s) used.

\begin{table}[ht]
\caption{In-domain and target domain performance of a YouTube-trained RNN-T, evaluated with and
  without fusion to a Voice Search LM (and normalizing YouTube LM in
  the case of the Density Ratio method).}
\label{tab:results_morphemes_vs}
\begin{center}
\begin{tabular}{|l|c|c|c|c|}
\hline
Model & WER & del & ins & sub  \\
\hline  \hline
YouTube $\rightarrow$ YouTube       & 11.3 & 2.4 & 1.9 & 7.0  \\
\hline
YouTube $\rightarrow$ Voice Search      & 17.5 & 3.9 & 4.1 & 9.6    \\
Shallow Fusion & 14.5 & 4.6 & 4.1 & 5.8    \\
Density Ratio $\lambda_\psi=\lambda_\tau$  & 13.0 & 3.3 & 3.2 & 6.5  \\
Density Ratio $\lambda_\psi, \lambda_\tau$  & {\bf 12.5} & 3.9 & 2.9 & 5.7 \\
\hline
\end{tabular}
\end{center}
\end{table}

\section{Fine-tuning a YouTube-trained RNN-T using limited Voice Search audio data}
\label{sec:yt_fine_tuning}

The experiments in Section~\ref{sec:yt_to_vs_expts} showed
that an LM trained on text from the target Voice Search
domain can boost the cross-domain performance of an RNN-T. The next
experiments examined fine-tuning the original YouTube-trained RNN-T on
varied, limited amounts of Voice Search \{audio, transcript\} data.
After fine-tuning, LM fusion was applied, again
comparing Shallow Fusion and the Density Ratio method.

Fine-tuning simply uses the YouTube-trained RNN-T model to warm-start
training on the limited Voice Search \{audio, transcript\} data.  This
is an effective way of leveraging the limited Voice Search audio data:
within a few thousand steps, the fine-tuned model reaches a decent
level of performance on the fine-tuning task -- though beyond that, it
over-trains. A held-out set can be used to gauge over-training and
stop training for varying amounts of fine-tuning data.

The experiments here fine-tuned the YouTube-trained RNN-T baseline
using 10 hours, 100 hours and 1000 hours of Voice Search data, as
described in Section~\ref{fig:training_data}. (The source
domain RNN-LM was not fine-tuned). For each fine-tuned
model, Shallow Fusion and the Density Ratio method were used to
evaluate incorporation of the Voice Search RNN-LM, described in
Section~\ref{sec:yt_to_vs_expts}, trained on text transcripts from the
much larger set of 21M Voice Search utterances.  As in
Section~\ref{sec:yt_to_vs_expts}, the dev set was used to tune the LM
scaling factor(s) and the sequence length scaling factor
$\beta$. To ease parameter tuning, the constraint $\lambda_\psi =
\lambda_\tau$ was used for the Density Ratio method.  The best
combinations of scaling factors from the dev set were then used to
generate the final eval results, which are shown in
Table~\ref{tab:results_morphemes_vs_ft}

\begin{table}[ht]
\caption{Fine tuning the YouTube-trained RNN-T baseline to the voice
  search target domain for different quantities of Voice Search
  fine-tuning data, evaluated with and without LM fusion on
  Voice Search test utterances. (Results for the ``no fine-tuning'' baseline
  carried over from Table~\ref{tab:results_morphemes_vs}).}
\label{tab:results_morphemes_vs_ft}
\begin{center}
\begin{tabular}{|l|c|c|c|}
\hline
Model                                             & WER & $\lambda$ & $\beta$ \\
\hline  \hline
Baseline (no fine-tuning)                        & 17.5         & - & -0.3 \\
Shallow Fusion                      & 14.5         & 0.3       & 0.6 \\
Density Ratio, $\lambda_\psi=\lambda_\tau$     & 13.0   & 0.5  & -0.3 \\
Density Ratio $\lambda_\psi, \lambda_\tau$  & {\bf 12.5} & 0.5, 0.6 & -0.1 \\
\hline
10h fine-tuning                           & 12.5        & - & 0.0 \\
Shallow Fusion          & 11.0        & 0.2 & 0.6 \\
Density Ratio, $\lambda_\psi=\lambda_\tau$   & 10.4 & 0.4 & 0.0 \\
Density Ratio, $\lambda_\psi,\lambda_\tau$   & {\bf 10.1} & 0.4, 0.45 & 0.0 \\
\hline
100h fine-tuning                           & 10.6        & - & 0.0 \\
Shallow Fusion          & 9.5       & 0.2 & 0.5 \\
Density Ratio, $\lambda_\psi=\lambda_\tau$   & {\bf 9.1} & 0.4 & 0.0 \\
\hline
1,000h fine-tuning                           & 9.5       & - & 0.0 \\
Shallow Fusion          & 8.8       & 0.2 & 0.5 \\
Density Ratio, $\lambda_\psi=\lambda_\tau$   & {\bf 8.5} & 0.3 & 0.0 \\
\hline
21,000h fine-tuning                           & 7.8       & - & 0.1 \\
Shallow Fusion          & 7.7       & 0.1 & 0.3 \\
Density Ratio, $\lambda_\psi=\lambda_\tau$   & {\bf 7.4} & 0.1 & 0.0 \\
\hline

\end{tabular}
\end{center}
\end{table}

\section{Discussion}

The experiments described here examined the generalization of a
YouTube-trained end-to-end RNN-T model to Voice Search speech data,
using varying quantities (from zero to 100\%) of Voice Search audio
data, and 100\% of the available Voice Search text data. The results
show that in spite of the vast range of acoustic and linguistic
patterns covered by the YouTube-trained model, it is still possible to
improve performance on Voice Search utterances significantly via Voice
Search specific fine-tuning and LM fusion. In particular, LM fusion
significantly boosts performance when only a limited quantity of Voice
Search fine-tuning data is used.

The Density Ratio method consistently outperformed Shallow Fusion for
the cross-domain scenarios examined, with and without fine-tuning to
audio data from the target domain. Furthermore, the gains in WER over
the baseline are significantly larger for the Density Ratio method
than for Shallow Fusion, with up to 28\% relative reduction in WER
(17.5\% $\rightarrow$ 12.5\%) compared to up to 17\% relative
reduction (17.5\% $\rightarrow$ 14.5\%) for Shallow Fusion, in the no
fine-tuning scenario.

Notably, the ``sweet spot'' of effective combinations
of LM scaling factor and sequence length scaling factor is
significantly larger for the Density Ratio method than for Shallow
Fusion (see Fig.~\ref{fig:heatmap_shallow} and
Fig.~\ref{fig:heatmap_ratio}).  Compared to Shallow Fusion, larger
absolute values of the scaling factor can be used.

A full sweep of the LM scaling factors ($\lambda_\psi$ and
$\lambda_\tau$) can improve over the constrained setting $\lambda_\psi
= \lambda_\tau$, though not by much.
Fig.~\ref{fig:heatmap_ratio_lambdas} shows that the optimal setting of
the two factors follows a roughly linear pattern along an off-diagonal
band.

Fine-tuning using transcribed Voice Search audio data leads to a large
boost in performance over the YouTube-trained baseline.
Nonetheless, both fusion methods give gains on top of fine-tuning,
especially for the limited quantities of fine-tuning data. With 10
hours of fine-tuning, the Density Ratio method gives a 20\% relative
gain in WER, compared to 12\% relative for Shallow Fusion.  For 1000
hours of fine-tuning data, the Density Ratio method gives a 10.5\%
relative gave over the fine-tuned baseline, compared to 7\% relative
for Shallow Fusion. Even for 21,000 hours of fine-tuning data,
i.e. the entire Voice Search training set, the Density Ratio method
gives an added boost, from 7.8\% to 7.4\% WER, a 5\% relative
improvement.

A clear weakness of the proposed method is the apparent need for
scaling factors on the LM outputs. In addition to the assumptions made
(outlined in Section~\ref{sec:deep_hybrid}), it is possible that this is due to
the implicit LM in the RNN-T being more limited than the RNN-LMs used.

\section{Summary}

This article proposed and evaluated experimentally an alternative to
Shallow Fusion for incorporation of an external LM into an end-to-end
RNN-T model applied to a target domain different from the source
domain it was trained on. The Density Ratio method is simple
conceptually, easy to implement, and grounded in Bayes' rule,
extending the classic hybrid ASR model to end-to-end models.  In
contrast, the most commonly reported approach to LM incorporation,
Shallow Fusion, has no clear interpretation from probability
theory. Evaluated on a YouTube $\rightarrow$ Voice Search cross-domain
scenario, the method was found to be effective, with up to 28\%
relative gains in word error over the non-fused baseline, and
consistently outperforming Shallow Fusion by a significant margin. The
method continues to produce gains when fine-tuning to paired target
domain data, though the gains diminish as more fine-tuning data is
used. Evaluation using a variety of cross-domain evaluation scenarios
is needed to establish the general effectiveness of the method.

\subsubsection*{Acknowledgments}

The authors thank Matt Shannon and Khe Chai Sim for valuable feedback
regarding this work.

\bibliographystyle{IEEEbib}
\bibliography{strings,refs}

\end{document}